\documentclass[reprint,aps,prl,twocolumn,notitlepage,superscriptaddress]{revtex4-1}
\usepackage{graphicx}
\usepackage{amsmath,amssymb,comment}
\usepackage{float}
\usepackage{comment,subfigure}
\usepackage{enumitem}
\usepackage[normalem]{ulem}
\usepackage[colorlinks=true,citecolor=blue,linkcolor=black]{hyperref}
\usepackage{graphicx}
\usepackage{bm,pgffor}
\usepackage{color}
\usepackage{dcolumn}
\usepackage{siunitx}
 
\usepackage{tikz}
\newcommand{\tikzcircle}[2][fill=white]{\tikz[baseline=-0.5ex]\draw[#1,radius=#2] (0,0) circle;}%

\begin{document}

\preprint{APS/123-QED}
\title{Absorbing-state transitions in granular materials close to jamming}
\date{\today}
\author{Christopher Ness}
\affiliation{Department of Chemical Engineering and Biotechnology, University of Cambridge, Cambridge CB3 0AS, United Kingdom}
\affiliation{School of Engineering, University of Edinburgh, Edinburgh EH9 3FB, United Kingdom}
\author{Michael E. Cates}
\affiliation{DAMTP, Centre for Mathematical Sciences, University of Cambridge, Cambridge CB3 0WA, United Kingdom}

\begin{abstract}
We consider a model for driven particulate matter in which absorbing states can be reached both by particle isolation and by particle caging.
The model predicts a non-equilibrium phase diagram in which analogues of hydrodynamic  and elastic reversibility emerge at low and high volume fractions respectively, partially separated by a diffusive, non-absorbing region.
We thus find a single phase boundary that spans the onset of chaos in sheared suspensions to the onset of yielding in jammed packings.
This boundary has the properties of a non-equilibrium second order phase transition,
leading us to write a Manna-like mean-field description that captures the model predictions.
Dependent on contact details, jamming marks either a direct transition between the two absorbing states, or occurs within the diffusive region.
\end{abstract}

\maketitle


Non-equilibrium phase transitions into absorbing states are of fundamental interest and relevant to applications such as spreading of infectious disease and reaction-diffusion problems~\cite{hinrichsen2000non,lubeck2004universal}.
Driven granular materials, both semi-dilute (volume fraction $\phi\approx0.1$) and jammed ($\phi>\phi_J\approx0.64$ (in 3D)), 
have proven to be useful experimental systems in which to study such transitions~\cite{corte2008random,nagamanasa2014experimental},
but the behaviour close to $\phi_J$ itself is unclear. 

In non-Brownian suspensions under oscillatory shear,
non-hydrodynamic particle contacts arise above a $\phi$-dependent critical strain amplitude $\gamma_c$,
moving the system from a Stokesian-reversible state to a chaotic, fluctuating one~\cite{pine2005chaos,corte2008random,corte2009self,hexner2015hyperuniformity,tjhung2015hyperuniform,pham2016origin}.
Meanwhile, jammed packings exhibit
a transition from elastic reversibility to plastic cage deformation at a $\gamma_c$ associated with yielding~\cite{maloney2006amorphous,fiocco2013oscillatory,regev2013onset,nagamanasa2014experimental,keim2014mechanical,leishangthem2017yielding,regev2015reversibility}.
The order parameter for both transitions may be chosen as the fraction $A$ of particles that are `alive',
that is, those whose position changes after successive shear cycles at steady state.
For $\gamma>\gamma_c$, time-irreversible particle contacts ($\phi<\phi_J$) and plastic rearrangements ($\phi>\phi_J$) render the systems active:
they have diffusion coefficient $\mathcal{D}>0$, with $A>0$ and all particles spending part of the time alive.
Below $\gamma_c$ the systems reach absorbing states with $A\simeq0$ and $\mathcal{D}=0$ due to hydrodynamic (elastic) reversibility when $\phi<\phi_J$ ($\phi>\phi_J$). 
In absorbing states most particles are never alive, but $A$ need not strictly vanish:
isolated per-cycle displacements are permitted provided the system is trapped in a finite basin of the phase space~\footnote{Although proving a useful diagnostic in the literature and the present work, $\mathcal{D}=0$ is not in general necessary for absorbing states, which may have a finite fraction of alive, diffusive particles that never interact with their dead counterparts.}.

The nonconserved order parameter $A$ carried by a conserved total number of particles, and the existence of multiple symmetry-unrelated absorbing states, should place these systems in the Manna class of non-equilibrium second order phase transitions~\cite{manna1991two,rossi2000universality,menon2009universality}.
This is borne out below $\phi_J$ in experiments~\cite{corte2009self}, molecular dynamics simulations~\cite{nagasawa2019classification} and in simplified models in which shear is mimicked by displacing overlapping particles~\cite{corte2009self,tjhung2015hyperuniform,hexner2017enhanced}. Above $\phi_J$, however, experimentalists have reported both second~\cite{nagamanasa2014experimental} and first~\cite{parmer2019strain} order behaviour, with simulations~\cite{das2019unified,nagasawa2019classification,parmer2019strain} consistently predicting the latter.
First order behaviour could be due to long-range elastoplastic effects present only above $\phi_J$ and not included in our model introduced below.
Setting this aside, and the differing governing forces on either side of $\phi_J$ (hydrodynamic \emph{vs.} elastic), the two transitions between absorbing and diffusive states share a number of features including
a diverging timescale for reaching the steady state~\cite{keim2013yielding,corte2008random,nagamanasa2014experimental,mobius2014ir}
and self-organisation~\cite{royer2015precisely,wang2018hyperuniformity} (including into hyperuniform states~\cite{weijs2015emergent,wu2015search,hexner2015hyperuniformity,tjhung2015hyperuniform}).
An important question thus emerges about whether, how and where the absorbing-state transitions that mark the absorbing-diffusive boundary meet as $\phi_J$ is approached from either side.

Recent computational studies of soft spheres under cyclic shear address this question~\cite{das2019unified,nagasawa2019classification},
revealing a convoluted non-equilibrium phase diagram whose interpretation within the context of absorbing-state transitions is hampered by complexities including point \emph{vs.} loop reversibility~\cite{schreck2013particle,lavrentovich2017period}, elasticity~\cite{boschan2016beyond}, and history dependence in $\phi_J$~\cite{das2019unified}. 
To make progress near $\phi_J$, simplified models building upon those of Refs~\cite{corte2009self,tjhung2015hyperuniform,hexner2017enhanced} are warranted.
Here we present such a model for driven particulate matter in which
particles cease to be alive
when they are either contact-free \emph{or} jammed.
The former serves as an analogue of hydrodynamic reversibility; the latter elastic reversibility. 
Our model predicts a non-equilibrium phase diagram exhibiting two distinct absorbing regions on either side of $\phi_J$ and an intermediate diffusive region.
The absorbing-diffusive transitions above and below $\phi_J$ show evidence of belonging to the Manna class,
while $\phi_J$ itself can, dependent on model parameters, mark a direct transition between absorbing regions,
or occur within the diffusive region.
A modified mean-field Manna description predicts the features of the phase diagram.

\paragraph{Model description:}

\begin{figure*}
\centering
\includegraphics[trim = 0mm 0mm 0mm 0mm, clip,width=1\textwidth,page=1]{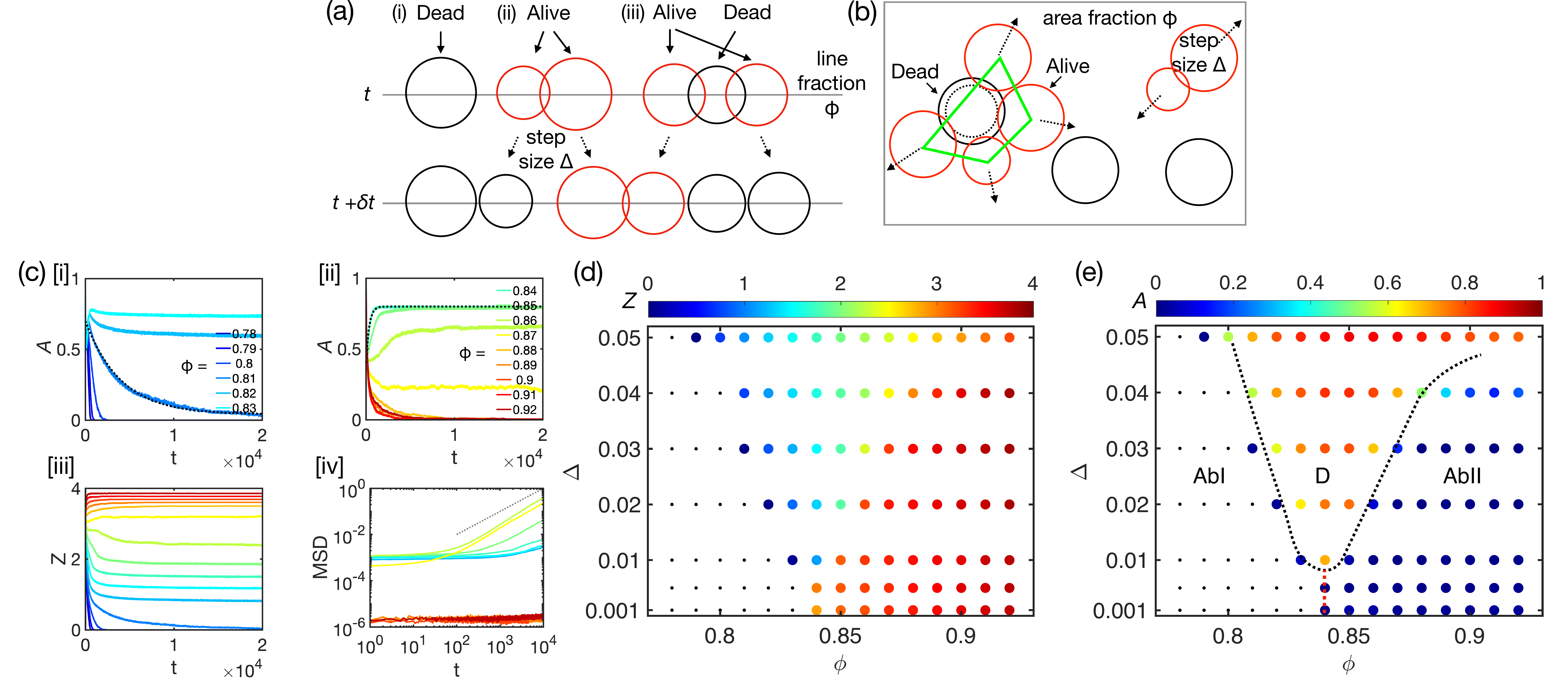}
\caption{
Model for driven particulate matter that combines particle isolation and caging.
Model definition in 1D (a) and 2D (b).
Contact-free particles and caged particles are dead (black circles \tikzcircle[black]{4pt}) and immobile;
all other particles are alive (red circles \tikzcircle[red]{4pt}) and move distance $\Delta$ along ${\bm n}_{ij}$ (dotted arrows) each timestep;
(c) Variation, with number of steps $t$, of:
[i]-[ii] $A$ (fraction of particles that are alive), with dotted lines showing example fits to $A = \alpha\exp(-t/\tau) + \kappa$;
[iii] average coordination number $Z$;
[iv] Mean Squared Displacement (grey line shows exponent 1, so that $\textrm{MSD} = \mathcal{D}t$);
 at different densities $\phi$ and step size $\Delta=0.03$.
 Colours in [iii]-[iv] follow the legends in [i]-[ii];
(d)-(e) Non-equilibrium phase diagrams showing (d) $Z$ and (e) $A$ as functions of density $\phi$ and step size~$\Delta$, measured after $N=2.5\times10^5$ steps;
dotted lines (black and red) are sketched to highlight phase boundaries,
black points represent values strictly 0.
}
\label{figure1}
\end{figure*}

The starting point for our model is a deterministic variant of that proposed by Tjhung \& Berthier (TB)~\cite{tjhung2015hyperuniform}, a member of the Manna class.
Random configurations of $N$ particles with radii normally distributed about $\sigma$ (variance $0.2\sigma$) are generated in a box of length $L$.
At each discrete timestep $t,t+\delta t,t+2\delta t,...$, we check for overlapping particles.
Particles with no overlapping neighbours (coordination number $z = 0$) are dead,
and are not moved from one timestep to the next.
Particles with $z>0$ (of which there are $N_A$) are alive and are displaced by
${\bm x}_i(t+\delta t) = {\bm x}_i(t) + \Delta\sum_{j = 1}^z {\bm n}_{ij}$,
where $\Delta$ is expressed in units of $\sigma$, and $\bm{n}_{ij}$ are unit vectors pointing to particle $i$ from each contacting particle $j$.
We write the mean squared displacement as $\mathrm{MSD}=\langle ({\bm x}_i(t) - {\bm x}_i(t_0))^2\rangle = \mathcal{D}t$, where $\langle \dots \rangle$ averages over particles.
The control parameters are the kick distance $\Delta$ and the density $\phi$, given by  $\sum_{i=1}^N2\sigma_i/L$ in 1D and $\sum_{i=1}^N \pi\sigma_i^2/L^2$ in 2D.
Cases that reach $A\equiv N_A/N\simeq0$ and $\mathcal{D}=0$ after some number of steps $t$ represent absorbing states,
while those with $A>0$, $D>0$ in steady state do not.
Notwithstanding differences in details, these dynamics produce a
non-equilibrium absorbing-diffusive phase transition of the same universality class as TB.

We next introduce a caging mechanism chosen so that jammed particles become dead,
describing first a 1D variant for simplicity. 
As above, particles with $z=0$ are stationary.
Particles that have $z=2$, i.e. contacts to their left and right,
are now also defined as dead
on the grounds that they are locally in a state of isostaticity.
A related constraint was imposed in a previous study of contact processes on a lattice~\cite{xu2013contact}.
Only particles with $z=1$ (or indeed $z>2$, which occurs at small $t$ due to the random initialisation) are alive and displaced at each $\delta t$ as above.
These dynamics are sketched in~Fig~\ref{figure1}(a).

For caging in 2D, our criterion stipulates the \emph{location} of contacting neighbours.
A particle is caged and therefore dead if its centre lies inside the polygon formed by connecting the centres of its overlapping neighbours,~Fig~\ref{figure1}(b).
This implies that a particle must have $z\geq3$ in order to be jammed, though this coordination alone is not necessarily sufficient.
An average coordination of $d+1$ represents the minimal value reachable in a packing of frictional particles~\cite{van2009jamming}.
To prevent large overlaps of caged particles,
we introduce a hard core
of radius $0.9\sigma_i$ for particle $i$, see dotted circle Fig~\ref{figure1}(b).
Any particle with a neighbour inside this hard core is declared alive, regardless of the arrangement of its other contacts.
In addition to caged particles, those with $z=0$ are dead~\footnote{Note that in our simplified models, which do not address intra-shear cycle dynamics, loop reversible states~\cite{schreck2013particle} are classified as dead.}.
All other particles are alive and are displaced at each $\delta t$ as above.
 We run the above dynamics for systems of $N= 5\times10^3$ particles~\footnote{Our results are consistent when measured with $10^3$ (following~\cite{corte2008random}) and $10^4$ particles, though more extensive measurement of the exponents and stringent tests for hyperuniformity will require larger systems (TB used $\mathcal{O}({10^5})$).}
 for $t = \mathcal{O}(10^5)$ timesteps,
 varying $\Delta$ and $\phi$ systematically and taking at least 30 realisations for each case.
 Discussed in the following are 2D results.

\paragraph{Model results:}
 
 Shown in Fig~\ref{figure1}(c) are plots of the evolution with $t$ of
the fraction of alive particles $A$,
the average coordination $Z \equiv \langle z\rangle$,
and the MSD,
for $\Delta = 0.03$ and $\phi=0.78-0.92$.
We find three distinct behaviours:

{\em \normalfont{Absorbing state I (AbI):}}
at $\phi$ $<0.81$, initially randomly positioned overlapping particles lose contact and $A$, $\mathcal{D}$ and $Z$ decrease with time, eventually reaching zero and marking entry into an absorbing state.
We identify this contact-free state as an analogue of the hydrodynamically reversible state reached in cyclicly sheared suspensions below $\gamma_c$~\cite{pine2005chaos}, 
in which there are no time-irreversible interactions in the system.
Note that period-multiplying is not observed in this region.

{\em \normalfont{Absorbing state II (AbII):}}
at $\phi>0.87$, $A$ decreases with time as particles form ubiquitous jammed cages
that result in $Z \geq3$ and $\mathcal{D}=0$ after long times.
In contrast to AbI, here $A$ does not reach zero but rather a steady value of order $10^{-3}$.
Snapshots of the simulation reveal rattler particles~\cite{majmudar2007jamming}
occupying vacancies in an otherwise stationary system,
indicating that the system is indeed in an absorbing state.
We identify this as the elastic reversibility region~\cite{fiocco2013oscillatory}.
Here there are time-irreversible interactions ({\em i.e.}, particle-particle contacts),
but their spatial arrangement leads to jamming at the per-particle level.
(Absorbing states of this kind cannot emerge under TB.)

{\em \normalfont{Diffusive (D):}}
for intermediate $\phi$, absorbing states are not reached, but rather after some transient the system reaches finite-$A$ steady states.
These are distinguished from those in AbII by the MSD (Fig~\ref{figure1}c[iv]):
clearly $\mathcal{D}>0$ here and the system is diffusive.

 At $\Delta=0.03$ our model thus predicts two types of absorbing state separated by a diffusive region, exhibiting an AbI-D-AbII sequence with increasing $\phi$.  AbI and AbII share $\mathcal{D}=0$ but are distinct in their mode of absorption: the former has $Z=0$, the latter $Z\geq3$. In the diffusive region D we find steady states with $0<Z<3$ and $A,\mathcal{D}>0$.

\paragraph{Non-equilibrium phase diagram:}

In Figs~\ref{figure1}(d) and (e) we present phase diagrams in the ($\phi$, $\Delta$) plane for $Z$ and $A$ respectively,
measured at steady state.
The $\Delta=0.03$ behaviour is retained for all $\Delta\geq0.01$,
with a broadening D region as $\Delta$ is increased.
For $\Delta <0.01$, we instead find that $Z$ increases over a narrow range of $\phi$,
with order parameter $A=\mathcal{O}(10^{-2})$ throughout.
This implies direct AbI-AbII jamming transitions (with no intermediate D region) at $\phi_J\approx0.84$ for $\Delta<0.01$ (red dotted line, Fig~\ref{figure1}(e)).
The properties and location of the AbI-D-AbII junction are examined further below.

\begin{figure}
\centering
\includegraphics[trim = 0mm 0mm 380mm 0mm, clip,width=0.5\textwidth,page=1]{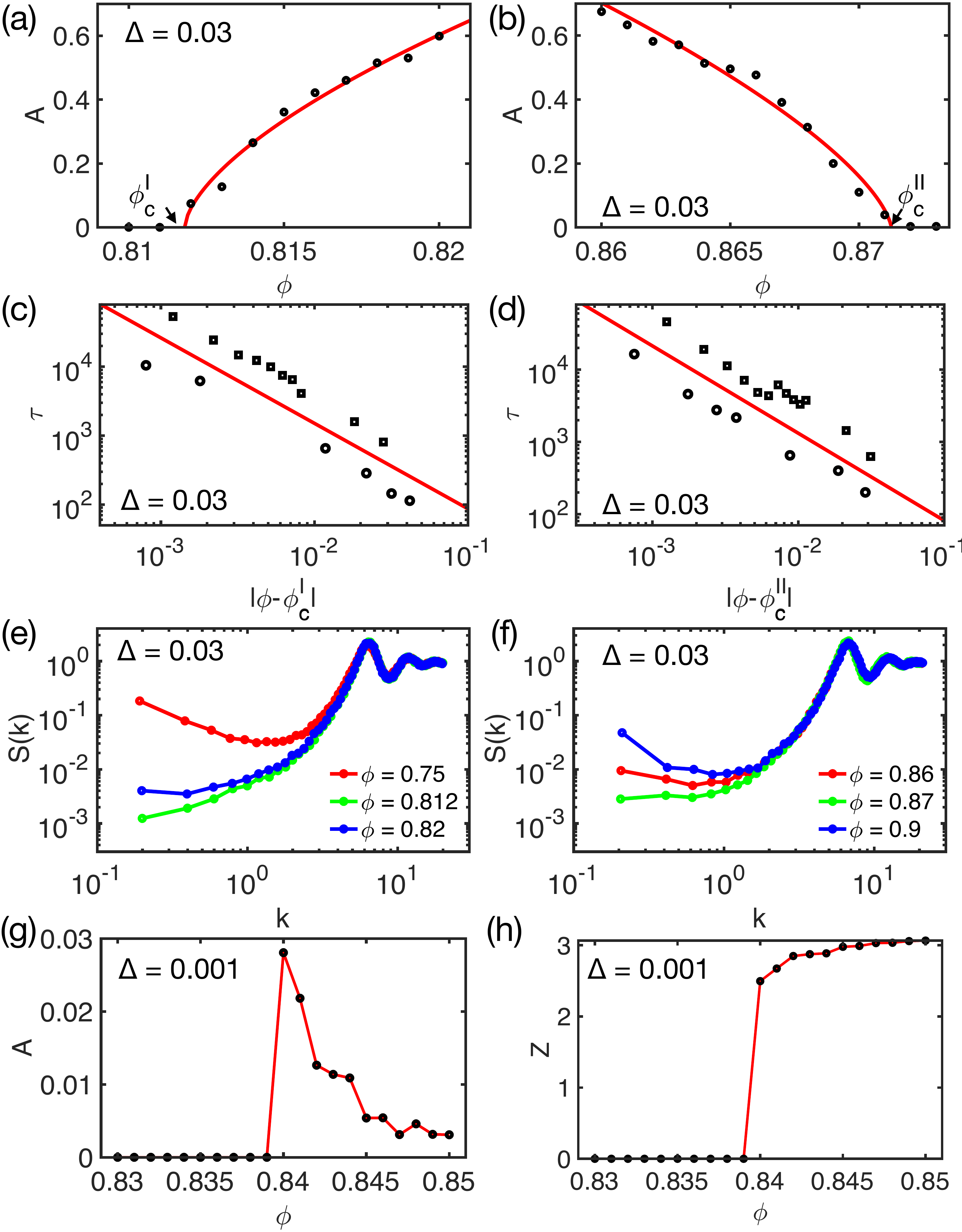}
\caption{
Properties of the phase boundaries.
(a), (b): The fraction of alive particles $A$ increases continuously from zero at $\phi_c^{I,II}$ with $\Delta=0.03$. Circles are model data; lines are fits to $A=k_1|\phi-\phi_c|^{\beta}$ with (a) $\phi_c^I=0.812$, $\beta^I=0.63$ and (b) $\phi_c^{II}=0.872$, $\beta^{II} = 0.67$.
The relaxation time $\tau$ diverges at (c) $\phi_c^I$ for AbI-D with exponent $\nu_\parallel=1.24$ and at (d) $\phi_c^{II}$ for D-AbII with $\nu_\parallel = 1.21$. Points are model data (circles are absorbing side of boundary; squares are diffusive side); lines are fits to $\tau = k_2|\phi-\phi_c|^{-\nu_\parallel}$.
Long wavelength density fluctuations are suppressed close to $\phi_c^I$ (e) and $\phi_c^{II}$ (f).
At the AbI-AbII boundary ($\Delta=0.001$), $A$ remains $\mathcal{O}(10^{-2})$ (g) and $Z$ increases discontinuously from 0 to $\approx2.5$ (h).
}
\label{figure2}
\end{figure}

We show next that the absorbing-diffusive boundary displays features of
a second order non-equilibrium phase transition on both sides of $\phi_J$.
Focussing again on $\Delta=0.03$, we find that close to the boundary (which occurs at $\phi_c(\Delta)$) $A$ can be written as $A=k_1|\phi-\phi_c|^{\beta}$ for both the AbI-D ($I$) and D-AbII ($II$) transitions, Fig~\ref{figure2}(a)-(b).
This is consistent with experiments~\cite{corte2008random,nagamanasa2014experimental} but not with the numerics of Ref~\cite{kawasaki2016macroscopic} that indicate a first order transition.
We find $\phi_c^I = 0.812$, $\phi_c^{II} = 0.872$, $\beta^I=0.63 \pm 0.02$ and $\beta^{II}=0.67\pm0.03$.
Following Ref~\cite{corte2008random} we write $A = \alpha\exp(-t/\tau) + \kappa$ (examples shown in dotted lines, Fig~\ref{figure1}(c)[i]-[ii]) leading to a relaxation time $\tau$ for the process.
$\tau$ diverges at $\phi_c^{I,II}$ according to $\tau = k_2|\phi-\phi_c|^{-\nu_\parallel}$, Fig~\ref{figure2}(c)-(d),
with $\nu_\parallel^I=1.24\pm0.04$ and $\nu_\parallel^{II}=1.21\pm0.03$.
Ref~\cite{lubeck2004universal} gives the exponents as $\beta=0.639$ and $\nu_\parallel=1.225$~\footnote{The 1D model variant similarly produces exponents consistent with Manna}.
We define a structure factor according to~\cite{berthier2011suppressed} 
$S(k) = \frac{1}{\phi L^2}\left(\left(\sum_i\sigma_i^2\cos({\bm x}_i\cdot {\bm k})\right)^2 +\left(\sum_i\sigma_i^2\sin({\bm x}_i\cdot {\bm k})\right)^2\right)$
for wave vector ${\bm k}$ and find that long wavelength density fluctuations are suppressed on the AbI-D and D-AbII boundaries, Fig~\ref{figure2}(e)-(f).
We defer a check for strict hyperuniformity~\cite{torquato2003local} (as done by TB for AbI-D) to future work.
Together these results show that the AbI-D and AbII-D transitions are consistent with the Manna class.
The direct AbI-AbII transition exhibits some similar features ($\tau$ diverges at $\phi_J$; $S(k)$ is suppressed at small $k$) but is distinct in that $A$ remains $\mathcal{O}(10^{-2})$, Fig~\ref{figure2}(g), and there is a discontinuity in Z (though no signature in the radial distribution function) implying first order behaviour, Fig~\ref{figure2}(h).

\paragraph{Mean-field description:}

\begin{figure}
\centering
\includegraphics[trim = 0mm 0mm 195mm 0mm, clip,width=0.5\textwidth,page=1]{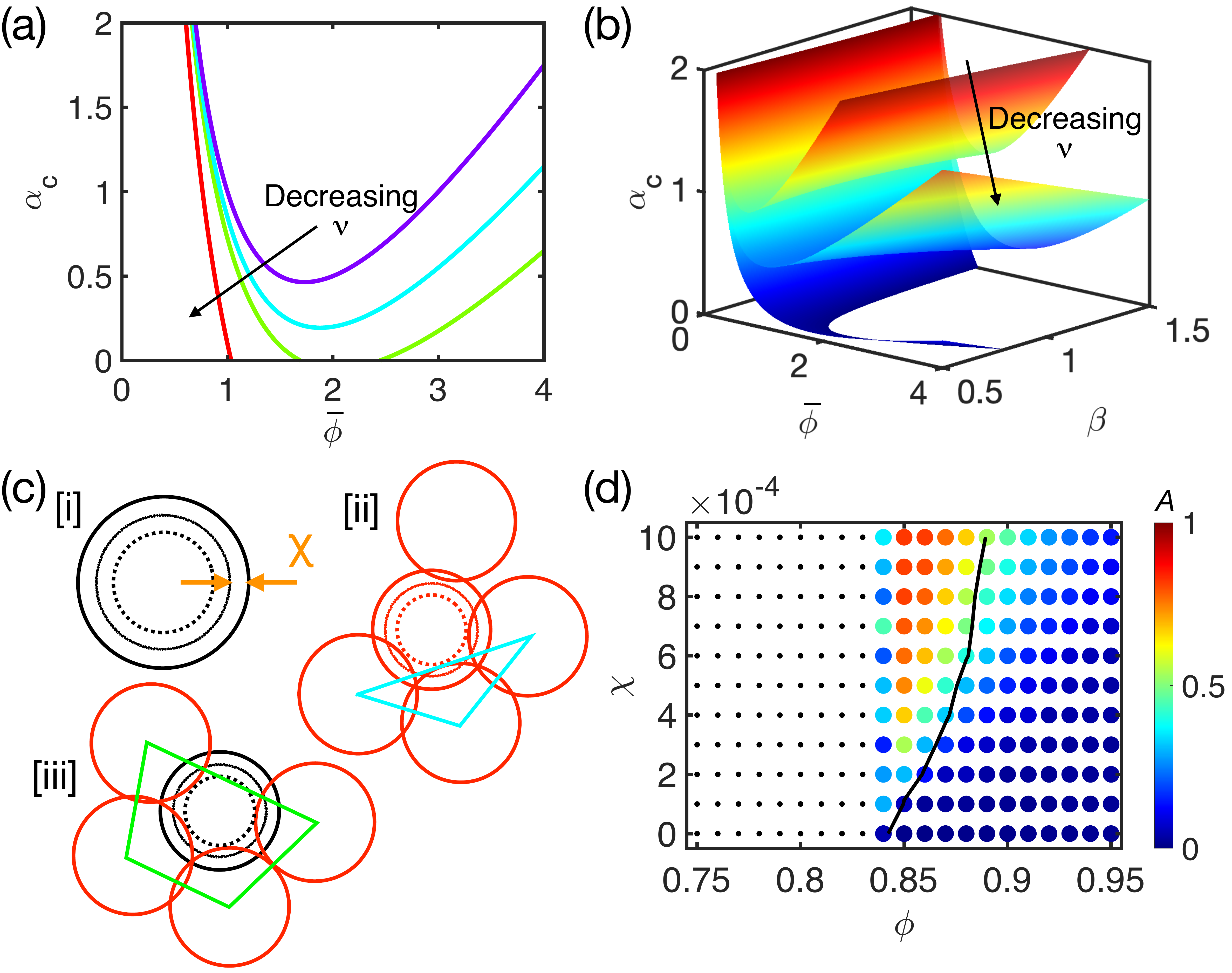}
\caption{
Predictions of modified mean-field Manna model, and a modified caging rule.
Shown are
(a) Eq~\ref{equation3} with $\beta=3$ and $\nu=[0.1, 0.725, 0.85, 1]$;
(b) Eq~\ref{equation3} with $\nu=[0.15, 0.5, 1]$;
(c) [i] Introducing a shell of width $\chi$: contacts lying within $\chi$ no longer contribute to caging; shown are [ii] unsuccessful cage; [iii] successful cage formation; 
(d) Increasing the thickness of the shell $\chi$ inhibits jamming, so that the AbI-AbII transition predicted for $\chi=0$ is replaced by a AbI-D-AbII sequence. 
Here $\Delta=0.001$ and the black line indicates the contour where $Z=3$, the minimum for system wide caging.}
\label{figure3}
\end{figure}

For a system of alive and dead particles
with densities $\rho_A$ and $\rho_B$ respectively, and $\phi=\rho_A + \rho_B$,
we write a modified mean-field Manna model (following~\cite{vespignani1997order,vespignani1998driving} and neglecting noise~\cite{hipke2009absorbing}) as:
\begin{equation}
\dot{\rho}_A =
{\nabla^2\rho_A}+
{\alpha\rho_A(\phi-\rho_A)}-
{ \beta\rho_A(1-\phi)}-
{\nu\rho_A\phi^2}
\end{equation}
with $\dot{\phi} = \nabla^2\rho_A$ and $\alpha, \beta, \nu > 0$.
Here $\alpha$ represents activation of dead particles by interaction with alive neighbours (related to $\Delta$ above),
$\beta$ represents isolated death
and $\nu$ accounts for death due to caging~\cite{xu2013contact}.
The quadratic $\phi$ dependence mimics our 1D model in which particles require 2 neighbours for death;
in 2D such a leading-order term is expected to emerge upon coarse graining even if not present initially, so we retain it in our minimal description.
Letting $\phi=\bar{\phi} + \psi$ and $\rho = \rho_A$, then letting
$\nabla^2 = \dot{\rho}=\dot{\psi}=\psi =0$,
leads to an expression for the critical driving rate
\begin{equation}
\alpha_c(\bar{\phi}) = \frac{\beta + \nu\bar{\phi}^2}{\bar{\phi}} - \beta
\label{equation3}
\end{equation}
beyond which the state at $\rho=0$ is linearly unstable to growth of activity.
This expression predicts a U-shaped boundary of critical driving rates, Fig~\ref{figure3}(a)-(b),
qualitatively consistent with the phase diagram predicted by our model, Fig~\ref{figure1}(e).
Importantly, Eq~\ref{equation3} predicts that the low-$\alpha$ extremum of the boundary 
can lie above, on or below the $\alpha=0$ axis, dependent upon the caging rate $\nu$.
When it lies above, a direct AbI-AbII jamming transition is expected. 
Otherwise the AbI and AbII states remain separated by a diffusive phase that includes the jamming point. 
This scenario implies that in the ($\phi$, $\Delta$) phase diagram, Fig~\ref{figure1}(e), the AbI-D-AbII junction should move to smaller $\Delta$ on decreasing the rate of caging. Alternatively, holding $\Delta$ fixed (below $0.01$) while inhibiting caging should cause the AbI-AbII transition to be replaced by a AbI-D-AbII sequence involving two phase transitions.

To test this idea we return to the simulation model and introduce a thin outer shell (width $\chi\sim10^{-4}\sigma$) to the particles.
We stipulate that caged death requires contacting neighbours to simultaneously form an enclosed polygon,
as above, and each have overlap distance $>\chi$, Fig~\ref{figure3}(c).
Now an isolated particle can be brought to life by any contact (as above),
whereas an alive particle can only achieve caged death
by contacts of sufficient overlap.
With this constraint, increasing $\chi$ should inhibit caged death and is thus expected to have the same effect on the phase diagram as decreasing $\nu$.

Indeed, we find that on fixing $\Delta=0.001$ and increasing $\chi$ we transit from the AbI-AbII behaviour in Fig~\ref{figure1}(e) to AbI-D-AbII behaviour, shown in Fig~\ref{figure3}(d).
Here the solid black line, clearly within the diffusive region for large $\chi$, marks points where $Z=3$.
While the exit from AbI appears to occur at a $\chi$-independent $\phi$, the additional requirement to exceed $\chi$ keeps particles alive so that larger $\phi$ must be reached to enter AbII.
As a result, the AbI-D-AbII junction in Fig~\ref{figure1}(e) passes downward through the $\Delta=0$ axis,
so that jamming, occurring at small $\Delta$,
 no longer marks a sharp absorbing-absorbing transition but instead occurs over a broadening range of $\phi$, within which the system is diffusive.
Together, our model and the mean-field expression suggest that jamming can manifest as the meeting of two distinct absorbing states (with $\phi_J$ dependent upon the governing dynamical rules), or may occur within the diffusive region.

\paragraph{Conclusion:}

We have presented a model of driven granular materials that places absorbing-state transitions in the vicinity of the jamming point within the Manna class.
The model predicts that the jamming point $\phi_J$ can either mark a transition between distinct types of absorbing state or can lie within a diffusive phase that separates such states.
An important open question is the extent to which more detailed aspects of this scenario are universal.
In particular, comprehensive reconciliation of recent findings~\cite{das2019unified,nagasawa2019classification} with the non-equilibrium phase diagram proposed here may require the inclusion in our model of spatially nonlocal effects such as elastoplasticity and long-range hydrodynamic interactions~\cite{boschan2016beyond}, perhaps guiding the development of new mean-field theories for amorphous materials.
Further work is warranted on the AbI-D and D-AbII boundaries 
 which correspond to conditions of maximal 
 data compressibility~\cite{martiniani2019quantifying}
 and mechanical memory storage~\cite{paulsen2014multiple,mukherji2019strength} respectively,
while fundamental understanding of driven transitions between contact-free, diffusive and jammed states
is relevant to suspension flow control~\cite{ness2018shaken} and soil liquefaction~\cite{huang2013review}.

\begin{acknowledgments}
We thank Elsen Tjhung, Cesare Nardini, Romain Mari and John Royer for useful discussions.  
Work funded in part by the European Research Council under the Horizon 2020 Programme, ERC grant agreement number 740269.
MEC is funded by the Royal Society.
CN acknowledges financial support from the Maudslay-Butler Research Fellowship at Pembroke College, Cambridge and latterly from the Royal Academy of Engineering under the Research Fellowship scheme. 
\end{acknowledgments}
\bibliography{library}

\end{document}